\documentclass[twocolumn,showpacs,preprintnumbers,amsmath,amssymb,aps,prb]{revtex4}
\usepackage{graphicx}
\begin{document}

\title{Dynamic Phases, Clustering, and Lane Formation for Driven Disk Systems 
in the Presence of Quenched Disorder} 
\author{
Y. Yang$^{1,2}$, D. McDermott$^{1,2}$, 
C. J. Olson Reichhardt$^{1}$, and  C. Reichhardt$^{1}$ 
} 
\affiliation{
$^1$Theoretical Division,
Los Alamos National Laboratory, Los Alamos, New Mexico 87545 USA\\ 
$^2$Department of Physics, Wabash College,
Crawfordsville, Indiana 47933, USA
} 

\date{\today}
\begin{abstract}

We numerically examine the
dynamic phases and pattern formation of
two-dimensional monodisperse repulsive disks
driven over random quenched disorder.
We show that there is a series of distinct dynamic regimes as  a function of increasing
drive, including a clogged or pile-up phase
near depinning, a homogeneous disordered flow state, and
a dynamically phase separated
regime consisting of
high density crystalline regions surrounded by a low density of
disordered disks.  At the highest drives
the disks arrange into one-dimensional moving lanes.
The phase separated regime has parallels with phase separation observed in
active matter systems, and arises in the disk system
due to the combination of nonequilibrium fluctuations
and density dependent mobility.
We discuss how this system exhibits pronounced differences
from previous studies of driven particles moving over random
substrates where the particles, such as superconducting
vortices or electron crystals, have longer range repulsive interactions,
and where dynamical phase separation and strong
one-dimensional moving chain effects are not observed.
The system we consider could be realized experimentally
using sterically interacting colloids driven over random pinning arrays or
quasi-two-dimensional granular matter flowing over rough landscapes.
\end{abstract}
\maketitle

\section{Introduction}

A wide range of systems can be
effectively modeled as a collection of
repulsively interacting particles
that are coupled to a substrate that serves as quenched disorder,
and these systems
typically exhibit a transition from a pinned to a sliding state 
under an applied external driving force \cite{1}.
Examples of such systems include vortices in
type-II superconductors \cite{2,3,4}, driven electron
or Wigner crystals \cite{5,6,7},
skyrmions in chiral magnets \cite{8,9}, charge stabilized colloids \cite{10,11,12},
and magnetically interacting colloidal systems \cite{13,14}.
The depinning transition
can either be elastic, where the particles keep their same neighbors, or plastic,
where the particles exchange neighbors and break apart \cite{1,3}.
In systems with intermediate or long range
repulsive particle-particle interactions, the ground state
is usually a defect-free triangular lattice.
When plastic depinning occurs, pinned and mobile particles coexist, 
leading to
a proliferation of topological defects in the
lattice and producing highly disordered particle configurations during plastic flow
\cite{1,2,3}.
At higher drives
there can be a transition from the plastic flow state to
a moving anisotropic crystal \cite{3,15,16} or
moving smectic state \cite{17,18,19,20}.
This translation is
associated with an increase in the  ordering of the system and
produces a distinct change in  the structure factor \cite{18,19,20} and
the density of topological defects \cite{18,20}
as well as cusps or dips in the transport curves and changes
in the fluctuation spectra \cite{20,21,22}.
Depending on the dimensionality and
anisotropy of the system, these dynamical transitions
can have continuous or first order characteristics \cite{1,3,23}.

In most of the systems where depinning and sliding dynamics
have been studied, the repulsive particle-particle interactions
are modeled as a  smooth potential that is either long range, as in the case of
Coulomb or logarithmic interactions, or screened long range, 
such as a Bessel function interaction for superconducting vortices or
a Yukawa interaction for colloidal systems.
There are many systems that have only short range hard disk type
particle-particle interactions,
such as granular matter or sterically interacting colloids.
Hard disk systems can exhibit very different behavior
than what is observed in systems with long
range repulsion, such as a strong density dependence
of the response near a crystallization or
jamming transition \cite{24,N}.
Two-dimensional (2D) systems
with long
range repulsive interactions form an
ordered solid down to very low densities
since the particles are always within interaction range of each other,
whereas
hard disk systems form a crystalline solid only for
the density at which the particles can just touch each other,
which corresponds to a packing density or area coverage of $\phi=0.9$ for
2D monodisperse nonfrictional disk packings \cite{24}.
For densities below the crystallization density, the
hard disk system forms a disordered or
liquidlike state.
It is not clear whether a hard disk assembly driven
over random disorder would exhibit the same
types of dynamical transitions
found for superconducting vortices, Wigner crystals, and colloids,
or whether it would simply form a moving disordered state at high drives.
Previous studies addressed
how pinning and obstacles affect the onset of the jamming transition in bidisperse
disk packs \cite{25,26}; however, the driven dynamics for nonzero loading
above the jammed states have not been studied.
Although it may seem that hard disks driven
over quenched disorder would
exhibit the same general dynamics, such as dynamical reordering at high drives,
as repulsive particle systems with longer range interactions,
the question has surprisingly
not previously been addressed.

Here we examine an assembly of monodisperse
harmonically interacting repulsive disks
driven over a random array of pinning sites.
We focus on disk densities $\phi < 0.9$,
below jamming or crystallization.  
Despite the simplicity of the model, we
find that this system exhibits a richer variety of dynamical phases
than those observed in studies of
longer range repulsive particles driven over
random disorder.
When the number of pinning sites is
smaller than the number of disks,
the pinned phase is associated with a pile up or clogging phenomenon
in which the system breaks up into clumps or clusters,
with unpinned disks prevented from moving by interactions with
disks trapped at pinning sites.
As the drive is increased beyond depinning,
the system enters either a fluctuating uniform disordered state or
a phase separated cluster state consisting of
a low density gas of disks coexisting with high density clusters.
Within the clusters, the disks form a predominantly triangular lattice.
The phase separated states generally appear when the driving
force is close to the value of the maximum pinning force.
For even higher drives, the system can transition
into a collection of one-dimensional (1D) moving lanes,
and the structure factor exhibits a strong smectic ordering signature.
We characterize the different phases and the transitions
between them using velocity-force curves,
the transverse root mean square displacements,
the structure factor, and the density of non-sixfold coordinated particles.

Dynamical phase separation does not normally occur
in systems with
longer range interactions
since the coexistence of a high density and a low density phase would have a
prohibitively large energy cost due to the longer range interactions in the dense phase.
For the disk system, the energy cost of the
particle-particle interactions is zero until the disks
come into contact, which occurs only at the highest densities.
Similarly, strong 1D laning occurs when the disks can approach each
other very closely in the direction of the applied drive without overlapping.
It is known that 2D granular systems that undergo
inelastic collisions can
exhibit cluster instabilities \cite{27,28}; however, in our
system there are no frictional contacts between the disks.
The density phase separated regime can be understood as
a type of active matter clustering effect,
where the combination of disk-disk collisions and pinning produce
nonequilibrium transverse fluctuations of the disks
as well as a density-dependent mobility.
Studies of active matter systems with short range particle-particle repulsion
and density-dependent mobility show similar clustering behavior \cite{29,30,31,32}.
At higher drives for the disk system, we find that a uniform moving state
forms when the transverse diffusion is lost.
Our work suggests that dynamical phase
separation and laning effects
are general features of driven systems with short range
hard disk particle-particle interactions moving over random disorder, which
could be realized experimentally by using sterically interacting colloidal
assemblies or quasi-2D granular matter flowing over random disorder.

\section{Simulation}
We consider a 2D system with
an area of $L^2$ with periodic boundary conditions in the $x$ and $y$-direction.
The sample contains $N_{d}$ harmonically repulsive disks of radius $R_{d}$
as well as $N_{p}$ pinning sites that
are modeled as non-overlapping parabolic
potential traps which can exert a maximum pinning force of $F_{p}$ on a disk.      
The disk dynamics are governed by the following overdamped equation of motion: 
\begin{equation}
\eta \frac{d {\bf R}_{i}}{dt} = {\bf F}_{dd}  + {\bf F}_{p}  + {\bf F}_{D} .
\end{equation}
Here $\eta$ is the damping constant 
and ${\bf R}_{i}$ is the location of disk $i$. 
The   
disk-disk interaction force is 
${\bf F}_{dd} = \sum_{i\neq j}k(2R_{d} - |{\bf r}_{ij}|)\Theta(2R_{d} - |{\bf r}_{ij}|) {\hat {\bf r}_{ij}}$,
where ${\bf r}_{ij} = {\bf R}_{i} - {\bf R}_{j}$,  $\hat {\bf r}_{ij}  = {\bf r}_{ij}/|{\bf r}_{ij}|$, and
the spring constant $k = 50$.
The pinning force ${\bf F}_p$
is modeled as arising from randomly placed parabolic attractive 
wells with a pinning radius of $r_{p} = 0.5$,
such that only a single disk can be trapped in a given pinning site at a time.
$F_p$ is the maximum force exerted by the pinning site at the edge of the well.
The driving force ${\bf F}_D=F_D{\bf \hat{x}}$ is applied along the $x$ direction,
and 
for each driving force we allow
at least $1 \times 10^6$ simulation time steps
to elapse before taking measurements
to ensure that the flow has reached a steady state.
At each value of $F_D$ we measure 
the average disk velocity
$\langle V_{x}\rangle = N_d^{-1}\sum^{N_d}_{i=1}{\bf v}_{i}\cdot {\hat{\bf x} }$,
where ${\bf v}_{i}$ is the
instantaneous velocity of disk $i$.
The density $\phi$ of the system is characterized by the
packing fraction or the area covered by the disks,
$\phi = N_{d}\pi R^2_{d}/L^2$, where $L = 50$ and $R_{d} = 0.5$
in dimensionless simulation length units.
In the absence of disorder, the disks form a polycrystalline state near
$\phi \approx 0.85$ 
and a triangular solid at $\phi \approx 0.9$.  
A variation of this model was previously used to study the depinning and jamming
of bidisperse disks driven over random pinning; in that work, with a disk radii ratio
of 1:1.4, the jamming density in a pin free sample was
$\phi_{j} \approx 0.845$ \cite{25}.    

\section{Varied Disk Density}

\begin{figure}
  \includegraphics[width=3.5in]{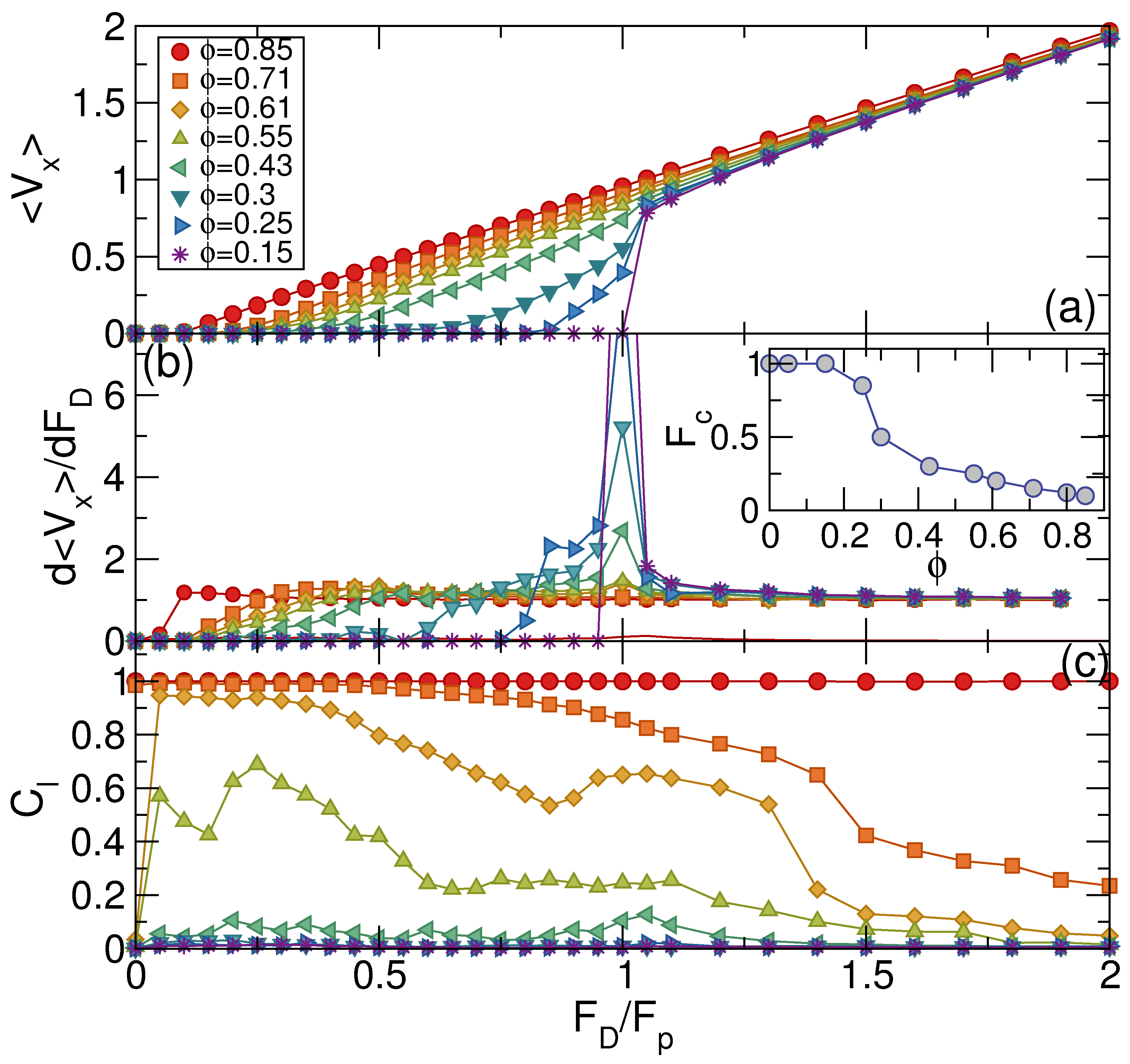}
\caption{
  (a) The average disk velocity $\langle V_{x}\rangle$ vs driving force
  $F_{D}/F_{p}$ for a system of harmonically interacting repulsive
  disks in a sample with $F_{p} = 1.0$
  and $N_{p} = 1440$ at disk densities of
  $\phi=0.85$ (red circles), $0.71$ (orange squares), $0.61$ (yellow diamonds),
  $0.55$ (light green up triangles), $0.43$ (medium green left triangles),
  $0.3$ (dark green down triangles), $0.25$ (blue right triangles), and  $0.15$ (purple stars).
  (b) The corresponding $d\langle V_x\rangle/dF_D$ vs $F_D/F_p$
  curves showing a peak near
  $F_D/F_p=1.0$.
Inset: The depinning threshold $F_{c}$ vs $\phi$, where 
$\phi \approx 0.3$ corresponds to a 1:1 ratio of disks to pinning sites.
(c) The corresponding cluster size $C_{L}$ vs $F_{D}/F_p$.
}
\label{fig:1}
\end{figure}

We first consider a fixed number of pinning sites
$N_p=1440$ with $F_{p} = 1.0$ as we vary the disk density from
$\phi=0.05$ to $\phi=0.85$, giving a ratio of
pinning sites to disks ranging from
$N_p/N_d=6.159$ to $N_p/N_d=0.37$.
With these parameters, a disk density of $\phi = 0.31$ corresponds
to a ratio of $N_p/N_d=1.0$.
Figure~\ref{fig:1}(a) shows $\langle V_{x}\rangle$ versus
$F_{D}/F_{p}$ for different values of $\phi$  and
Fig.~\ref{fig:1}(b) shows the corresponding $d\langle V_{x}\rangle/dF_{D}$ curves.
In the inset of Fig.~\ref{fig:1}(b) we plot
the depinning force $F_{c}$ vs $\phi$ indicating that
$F_{c}$ has a constant value of $F_c \approx F_p$ at low disk densities
$N_{p}/N_{d} > 5.0$.
In this density range,
almost every disk can be pinned directly by a pinning site,
so collective interactions between the disks do not play an important role in
the depinning process; instead, depinning occurs in the single particle limit and
the depinning threshold is determined only by the value of
$F_{p}$.
For $N_{p}/N_{d} < 1.0$, some of the disks are
not trapped by pinning sites, and these untrapped disks exert a force on the
pinned disks 
which lowers the depinning threshold, as shown in the inset of Fig.~\ref{fig:1}(a).

In Fig.~\ref{fig:1}(b),
for  $\phi \leq 0.55$
there is a pronounced peak in
$d\langle V_{x}\rangle/dF _{D}$
near $F_{D}/F_{p} = 1.0$.  This corresponds to the maximum pinning force
from the substrate, so that for $F_{D}/F_{p} >1.0$ all the disks are moving.
For
$N_{p}/N_{d} > 0.8$
or $\phi < 0.4$, a large fraction of the disks
are located at pinning sites
and the collision rate is low, so that most of the disks do not become mobile
until $F_{D}/F_{p} > 1.0$, producing the jump in $\langle V_{x}\rangle$
at depinning at the lower fillings.
For $N_{p}/N_{d} < 1.0$, there are excess disks that cannot be trapped
directly by the pinning sites, and in principle these disks would be mobile
for arbitrarily low $F_{D}$;
however, they can still be indirectly pinned
or blocked by  disks that are located at the pinning sites, creating a local pile up
or clogging configuration \cite{25}.
Since these interstitial disks exert forces on the disks located at the
pinning sites, their presence reduces the depinning threshold
by more than a factor of 2.
For fillings
$N_p/N_{d}=1.0$ to $0.571$,
corresponding to $0.3 \leq \phi \leq 0.55$, some disks remain
pinned until $F_{D} \geq F_{p}$, producing
a weak peak in the $d\langle V_{x}\rangle/dF_{D}$ curves
at $F_D/F_p=1.0$. When $\phi$ is large enough, most of the disks are already
moving for $F_{D}/F_{p} < 1.0$, and the peak feature is lost.

In Fig.~\ref{fig:1}(c) we plot the average value $C_l$
of the size of the largest cluster normalized by the number of disks
in the system as a function of $F_{D}/F_{p}$.
To determine $C_l$, we use the cluster counting
algorithm of Luding and Herrmann \cite{33}.
For $\phi < 0.43$, $C_{l}$ is low and the largest clusters contain 10 or fewer disks.
For $\phi \geq 0.43$, there is an increase in the cluster size at low drives
due to a pile up effect in which unpinned disks
accumulate behind pinned disks.
For $\phi = 0.85$, the system forms a large cluster and $C_{l} = 1.0$ for all $F_{D}$.
At $\phi  = 0.55$, 0.61, and $0.71$, there is a drop off in $C_{l}$
for $F_{D}/F_p > 1.05$, $1.33$, and $1.4$, respectively,
indicating a decrease in the cluster size.
There is also a local maximum in $C_l$ near $F_{D}/F_{p} = 1.0$
at $\phi=0.61$.

\subsection{Intermediate Disk Densities}

\begin{figure}
  \includegraphics[width=3.5in]{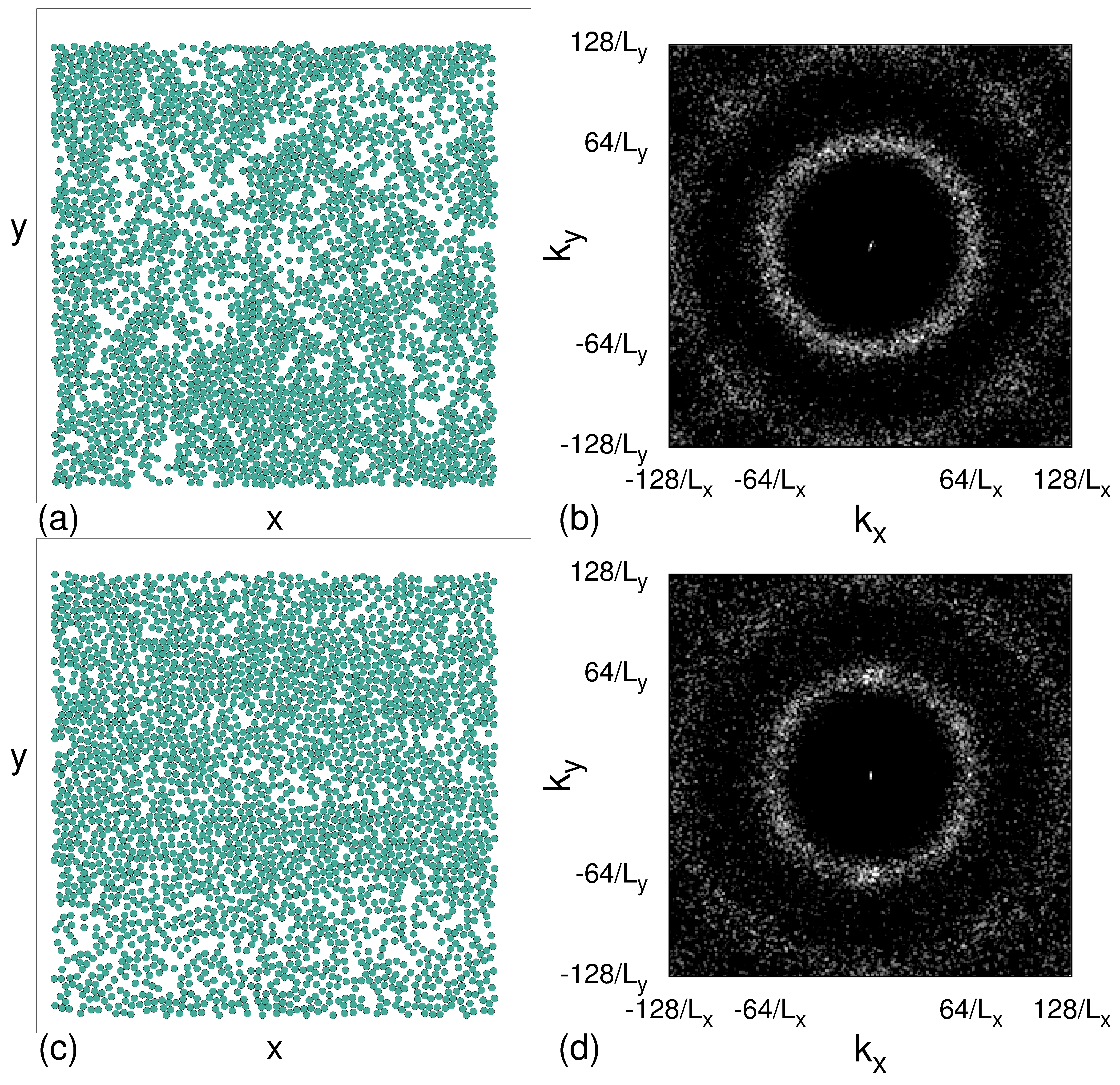}
\caption{(a) The disk positions (circles) for the system in
  Fig.~\ref{fig:1} at $\phi = 0.61$ for $F_{D}/F_{p} = 0.3$, showing
  a clustering or pile up effect.
  (b) The corresponding structure factor $S({\bf k})$ has a ringlike signature.
  (c) The  driven homogeneous phase
  in the same system at $F_{D}/F_{p} = 0.7$.
  (d) The corresponding $S({\bf k})$ plot from (c).
}
\label{fig:2}
\end{figure}

In Fig.~\ref{fig:1}(c), for $\phi = 0.61$ there is an initial increase in
$C_{l}$ up to
$C_l=0.95$ at small but finite $F_{D}/F_p$ due to the pile up effect.
This is followed by a decrease in $C_l$ to a local minimum near $F_{D}/F_{p} = 0.85$,
and then by another increase to a
local maximum in the range $0.85  < F_{D}/F_{p} < 1.4$,
indicating a growth in the size of the largest cluster
near $F_{D}/F_p = 1.0$.
In Fig.~\ref{fig:2}(a) we plot the disk configurations for the $\phi = 0.61$ system at
$F_{D}/F_{p} = 0.3$ where $C_{l} = 0.95$ showing large scale clustering.
Similar configurations appear at
$F_{D}/F_{p} = 0.3$ for $0.43 < \phi <  0.85$.
In Fig.~\ref{fig:2}(b), the corresponding structure factor
$S({\bf k})=N_d^{-1}|\sum_i^{N_d}\exp(-i{\bf k} \cdot {\bf r}_i)|^2$
of the disk configuration has
a ringlike feature indicative of a disordered system.
As the drive is increased beyond
the depinning transition, the clusters break apart and the
disk  density becomes homogeneous,
as shown in Fig.~\ref{fig:2}(c) for $F_{D}/F_{p} = 0.7$, where
a reduction in $C_l$ has occurred.
The corresponding structure factor in Fig.~\ref{fig:2}(d) 
still contains a ringlike feature but has excess weight in two peaks along $k_{x}  = 0$,
indicating the formation of some chainlike structures due to the
$x$-direction driving.

\begin{figure}
  \includegraphics[width=3.5in]{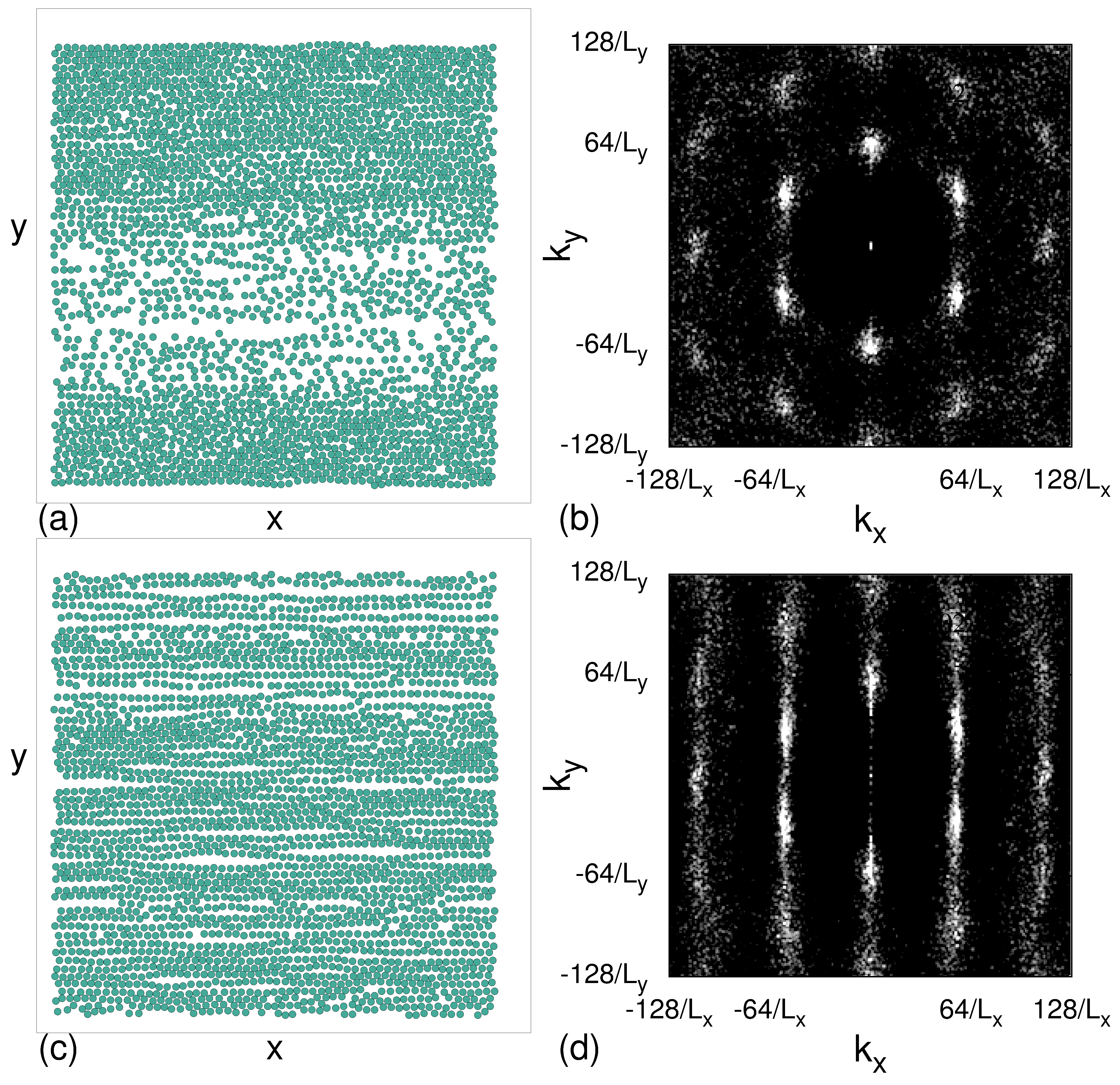}
\caption{ The disk positions (circles) for the system in Fig.~\ref{fig:1}
  at $\phi = 0.61$ for $F_{D}/F_p = 1.05$, corresponding to the
  local maximum in $C_{l}$ in Fig.~\ref{fig:1}(c).
  Here the system forms a density phase separated state.
  (b) The corresponding $S({\bf k})$ plot contains
 sixfold peaks due to the triangular ordering in the dense phase.
 (c) The same system at $F_{D}/F_p = 2.0$ where a moving chainlike state
 forms. (d) The corresponding $S({\bf k})$ shows smectic ordering.
}
\label{fig:3}
\end{figure}

For $0.7 < F_{D}/F_p <  1.4$, the system forms a density phase separated state,
as illustrated in Fig.~\ref{fig:3}(a)
for $F_{D}/F_p = 1.05$.  Here
there is a high density region with $\phi \approx 0.85$ in which
the disks have triangular ordering coexisting with
a low density region where the disks are disordered.
The corresponding structure factor in Fig.~\ref{fig:3}(b) shows six
peaks due to the triangular ordering within the dense phase.
There is some smearing of the peaks along $k_{y}$ due to the tendency of the
crystallites in the dense phase to align with the driving direction.
For $F_{D}/F_p > 1.4$, where $C_{l}$ drops, the disks become more spread out and form
1D moving chains of the type shown in
Fig.~\ref{fig:3}(c) at $F_{D}/F_p = 2.0$.
The corresponding $S({\bf k})$ in Fig.~\ref{fig:3}(d) has strong smectic ordering.
In general,
for $\phi \geq 0.43$
we find a phase separation in the vicinity of $F_D/F_p \approx 1$ similar to that
shown in Fig.~\ref{fig:3}(a), 
where the extent of the dense region grows with increasing
$\phi$ while the low density regions become smaller.

\subsection{Low Disk Density}

\begin{figure}
  \includegraphics[width=3.5in]{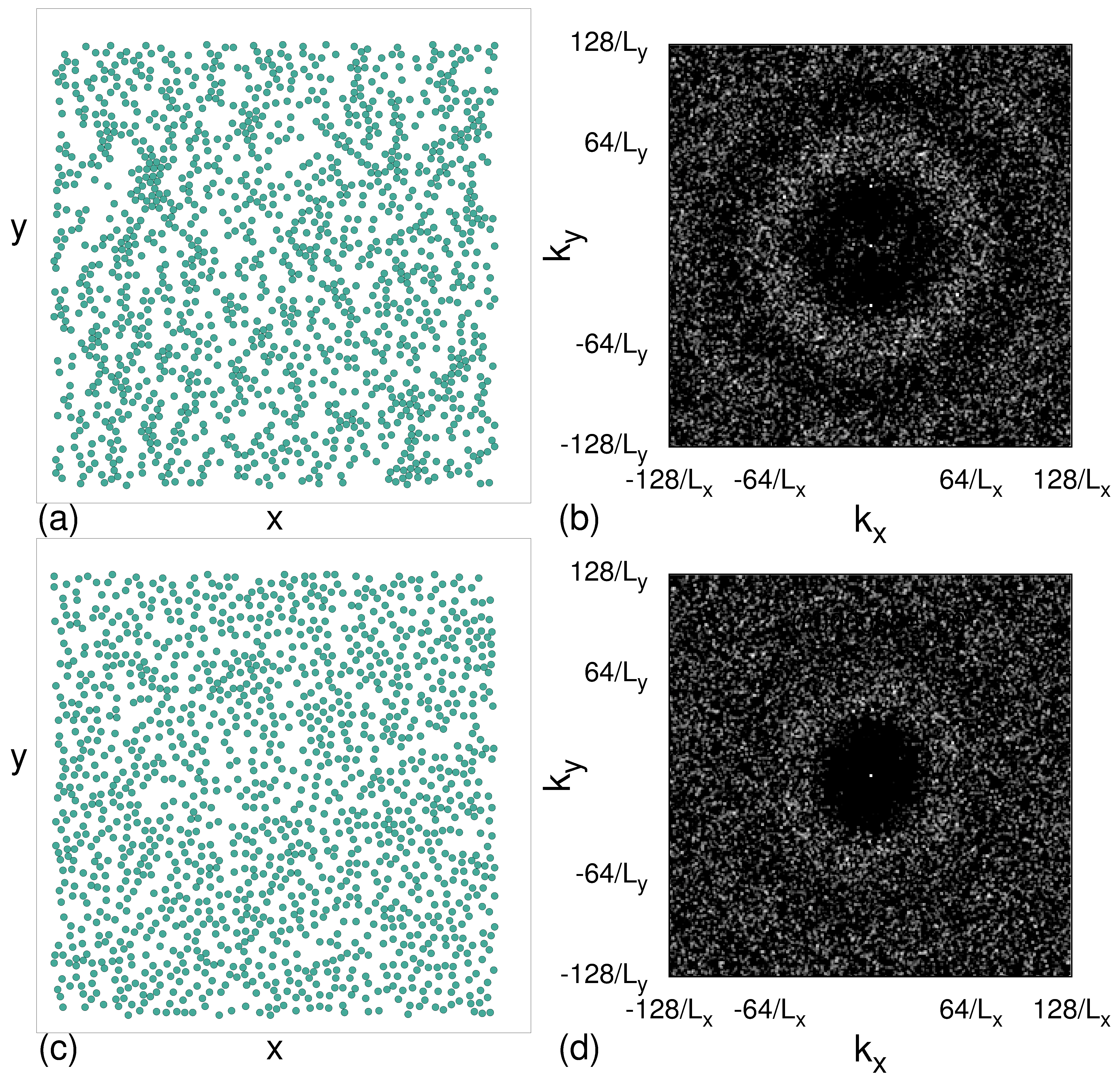}
\caption{
  (a) The disk positions (circles) for the system in Fig.~\ref{fig:1} at $\phi = 0.3$
  for $F_{D}/F_{p} = 0.15$, showing the formation of small clusters.
(b) The corresponding $S({\bf k})$ plot.
  (c) The same system at $F_{D}/F_{p} = 0.6$
  in the moving phase where the disk density becomes homogeneous.
  (d) The corresponding $S({\bf k})$ shows a diffuse or liquidlike pattern.}
\label{fig:4}
\end{figure}

\begin{figure}
  \includegraphics[width=3.5in]{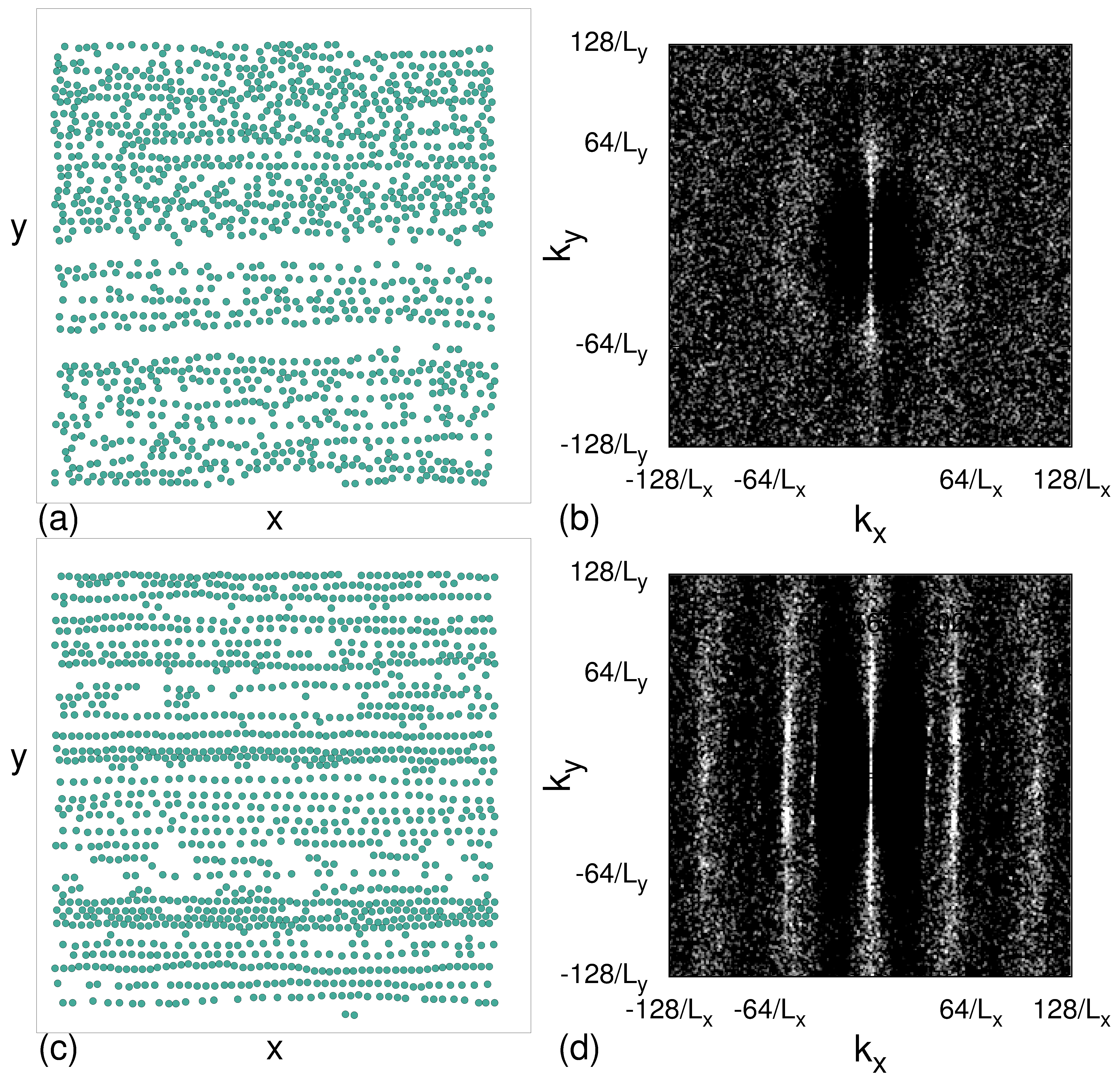}
\caption{
  (a) The disk positions (circles) for the system in
  Fig.~\ref{fig:1} at $\phi = 0.3$ for $F_{D}/F_{p} = 1.05$, where the disks form
chainlike patterns.
(b) The corresponding $S({\bf k})$ plot.
(c) The same system at $F_{D}/F_{p} = 2.0$ in the moving phase
where the disks form a series of chains or stripes. (d) The corresponding $S({\bf k})$
has smectic ordering.
}
\label{fig:5}
\end{figure}

For $\phi < 0.43$, the clumps that form near depinning are small, as illustrated
in Fig.~\ref{fig:4}(a) at $\phi = 0.3$ and $F_{D}/F_{p} = 0.15$.
The clumps are anisotropic and 
show some alignment along the $y$-direction,
while the corresponding structure factor in Fig.~\ref{fig:4}(b)
has a ringlike signature.
At higher drives above depinning when some of the disks are moving,
the disk density is more homogeneous, as
shown in Fig.~\ref{fig:4}(c) at $F_{D}/F_{p} = 0.6$.
The corresponding $S({\bf k})$ plot
in Fig.~\ref{fig:4}(d) has a more diffuse structure.
Near $F_{D}/F_{p} = 1.0$, most of the disks
are in motion and form chainlike structures, as illustrated in
Fig.~\ref{fig:5}(a,b) for $F_{D}/F_p = 1.05$.
The disk density is not uniform, with some chains closer together and others
further apart; however, the denser regions are still too sparse to form sections
of triangular lattice of the type that appear at $\phi=0.61$ in Fig.~\ref{fig:3}(a).
As $F_{D}$ increases for the $\phi=0.3$ sample, the
moving chains of disks become better defined, as shown
in Fig.~\ref{fig:5}(c) at $F_{D}/F_p = 2.0$.
The interchain spacing becomes small enough that the disks in neighboring
chains are almost touching,
and the corresponding
structure factor in Fig.~\ref{fig:5}(d) shows strong smearing along the $k_{y}$ direction.

These results indicate that even though $\phi$ is below the close-packed density of
$\phi=0.9$,
different dynamic phases can arise and there can be
transitions into states with smectic ordering,
similar to the smectic states observed for driven superconducting vortices
\cite{7,18,19,20,22}.
In general, the 1D channeling effect illustrated in Fig.~\ref{fig:5}(c)
is much more pronounced in the disk system
than in systems with longer range interactions.
The moving disks are unstable against the formation of chainlike structures due to
a velocity collapse phenomenon.  If one moving disk slows down, the disk immediately
behind it can run into it and cause it to speed up again, but once the two disks move
beyond their steric interaction range, there are no particle-particle interactions to push them
further apart, so the disks tend to pile up behind each other in the longitudinal
direction.
For $\phi = 0.85$, the system forms a dense cluster with polycrystalline
triangular ordering, and for $F_{D}/F_{p} > 1.0$ the disks
form a single triangular domain that is aligned with the driving direction.

\subsection{Transverse Diffusion and Topological Order}

\begin{figure}
  \includegraphics[width=3.5in]{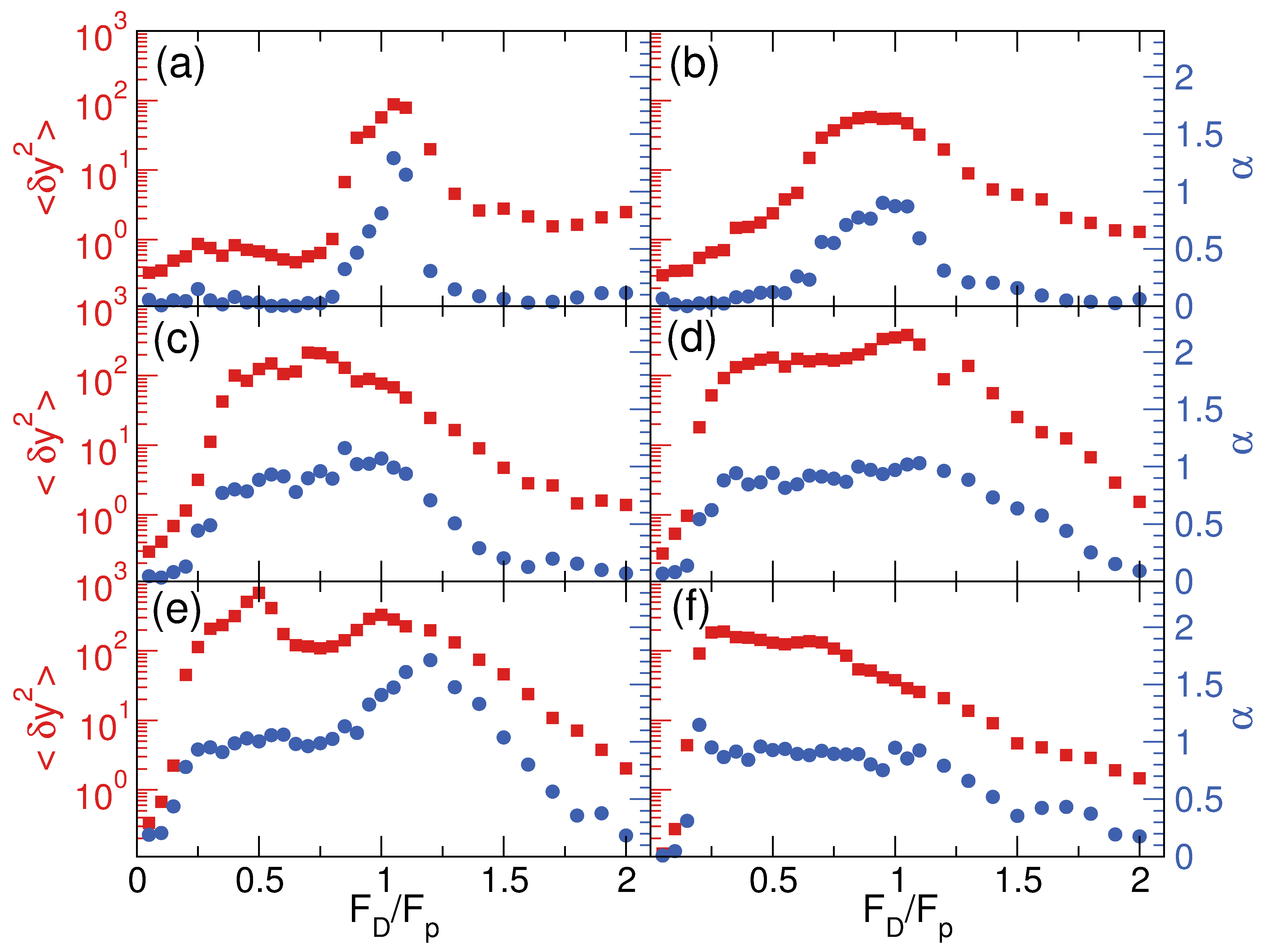}
\caption{
  The transverse displacements $\langle \delta y^2\rangle$
  obtained after $4\times 10^6$ simulation time steps (red squares)
  and the diffusive exponent $\alpha$ (blue circles) vs $F_D/F_p$
  for the system in Fig.~\ref{fig:1} at $\phi=$
   (a) 0.25, (b) $0.3$, (c) $0.43$, (d) $0.55$, (e) $0.61$, and (f) $0.71$.
}
\label{fig:6}
\end{figure}

We can characterize the different phases by measuring the particle displacements in the
direction transverse to the applied drive,
$\langle \delta y^2\rangle = N_d^{-1}\sum_{i=1}^{N_d}(y_i(t) - y_i(t_{0}))^2$,
for varied $F_{D}/F_{p}$.
In general we find
$\langle \delta y^2\rangle \propto t^{\alpha}$ at long times.
In the disordered homogeneous density regimes, $\alpha=1.0$,
indicative of diffusive behavior, while
$\alpha < 1.0$ just above depinning and in the moving chain state.
In Fig.~\ref{fig:6} we plot the value of $\langle \delta y^2\rangle$ obtained at a
fixed time of $5\times10^6$ simulation time steps versus $F_D/F_p$ along with
the corresponding 
value of $\alpha$ for the system in Fig.~\ref{fig:1}
at $\phi = 0.25$, 0.3, 0.43, 0.55, 0.61, and $0.71$.
For $\phi = 0.25$ and $\phi=0.3$ in Fig.~\ref{fig:6}(a,b),
there is a peak in $\langle \delta y^2\rangle$ near $F_{D}/F_{p} = 1.0$, where
$\alpha \approx 1.0$, indicating diffusive behavior.
The maximum amount of transverse diffusion falls at the same
value of $F_D/F_p$ as the  peak in
$d\langle V_{x}\rangle/dF_{D}$ shown in Fig.~\ref{fig:1}(b).
At low drives where the system forms a clogged state,
the transverse diffusion is suppressed.
At higher drives where the disks form 1D channels,
the diffusion in the direction transverse to the drive is
strongly suppressed and $\alpha \rightarrow 0$,
indicating that the 1D channels
are frozen in the transverse direction.

For $ \phi = 0.43$, 0.55, and $0.61$ in Fig.~\ref{fig:6}(c,d,e),
$\langle \delta y^2\rangle$ has a
double peak feature.
The first peak corresponds to the onset of the
homogeneous moving phase,
while the second peak occurs when
the system starts to undergo phase separation.
For $\phi = 0.61$, where the strongest phase separation is observed,
there is even a region of drive for which
$\langle \delta_y^2\rangle$ exhibits superdiffusive
behavior with  $\alpha > 1.0$.
At longer times the behavior transitions to regular diffusion.
For higher drives, both $\langle \delta_y^2\rangle$ and $\alpha$ decrease
with increasing drive
as the system forms a moving chain state.
For $\phi = 0.71$ in Fig.~\ref{fig:6}(f), the double peak feature begins to
disappear.
Numerical studies of vortices in type-II superconductors \cite{22,34}
show that the vortices exhibit strong transverse diffusion above the depinning
transition,
while at higher drives where a moving smectic state appears,
the transverse diffusion is strongly suppressed and the
system freezes in the transverse direction.
The vortex system typically has only 
a single peak in $\langle \delta y^2\rangle$
rather than the double peaks we observe here.
The regime
of superdiffusive behavior for $\phi = 0.61$
arises due to collective transverse motion of the disks in the dense phase.

Another measure often used to characterize
interacting particles driven over disorder is the fraction
$P_6$ of sixfold coordinated particles.
Here $P_6=N_d^{-1}\sum_{i}^{N_d}\delta(z_i-6)$, where $z_i$ is the coordination
number of disk $i$ obtained from a Voronoi tesselation.
In the case of superconducting vortices in the absence of
pinning, the ground state is a triangular lattice with $P_{6} = 1.0$,
while when
strong disorder is present, the pinned state 
is disordered and contains numerous topological defects so that $P_{6} < 1.0$.
At high drives, where the effect of pinning is reduced, the
system can dynamically reorder into a moving triangular lattice
with $P_{6} = 1.0$ or into a moving smectic where
some topological defects persist that are aligned with the direction of drive,
giving $P_6 \lesssim 1$ 
\cite{3,4,15,18,19,20,22}.

\begin{figure}
  \includegraphics[width=3.5in]{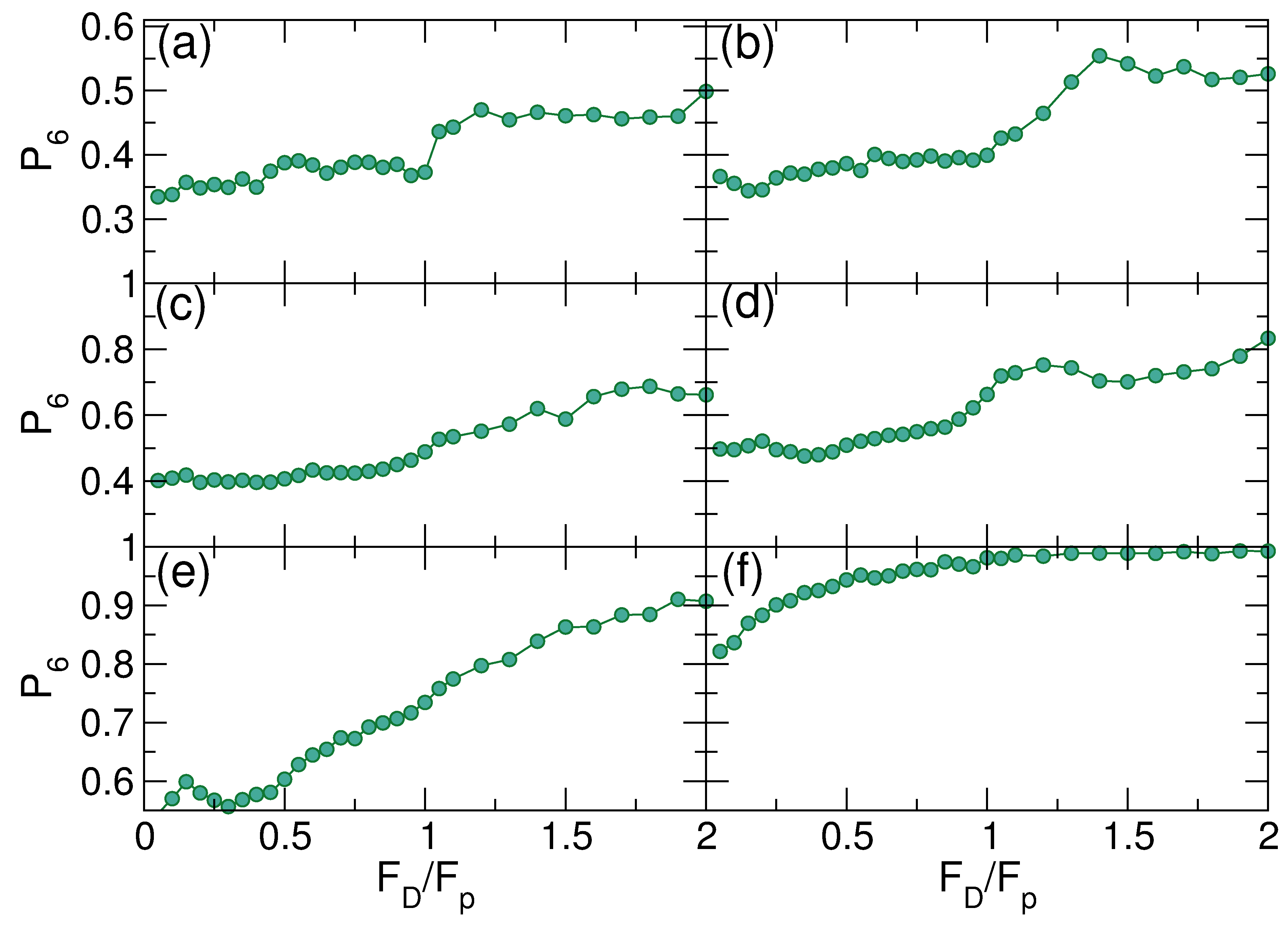}
\caption{ The fraction $P_6$ of sixfold coordinated disks vs $F_{D}/F_{p}$ for the system in
  Fig.~\ref{fig:1} for $\phi =$ (a) 0.25, (b) $0.3$, (c) $0.43$, (d) $0.61$,
  (e) $0.71$, and (f) $0.85$.
  For $\phi = 0.61$ in panel (d), the local maximum in $P_{6}$
  near $F_{D} = 1.0$ is correlated with the formation of the phase separated
state shown in Fig.~\ref{fig:3}(a).
}
\label{fig:7}
\end{figure}

In Fig.~\ref{fig:7} we plot $P_{6}$
versus $F_{D}/F_{p}$
for the system in Fig.~\ref{fig:1} at $\phi = 0.25$, 0.3, 0.43, 0.61, 0.71, and $0.85$.
Although there are several similarities to the behavior of $P_6$ observed for
superconducting vortices,
there are a number of notable differences.
For $\phi=0.25$ and $\phi=0.3$ in Fig.~\ref{fig:7}(a,b), there is an
increase in $P_{6}$ above $F_{D}/F_{p}=1.0$ which corresponds to the formation of the
moving  chain state illustrated in Fig.~\ref{fig:5}(a), followed by a saturation of
$P_6$ at higher drives
to $P_{6} = 0.55$.
This is in marked contrast to the behavior observed in the vortex system,
where $P_{6}$ saturates to a value much closer to $P_6=1.0$
due to the longer range particle-particle repulsion
which favors the formation of a triangular vortex lattice
down to quite low vortex densities.
At $\phi = 0.43$ in Fig.~\ref{fig:7}(c),
$P_6$ shows a similar trend as in the systems with lower disk densities;
however, $P_{6}$ saturates
to a higher value of $P_6=0.68$.
In Fig.~\ref{fig:7}(d) at
$\phi = 0.61$, there is a local maximum in $P_{6}$ for
$0.9 < F_{D}/F_p < 1.4$ that coincides with the density phase separated regime.
The disks in the dense phase have mostly triangular ordering,
as shown in Fig.~\ref{fig:3}(a,b).
For higher drives of $F_{D}/F_{p} > 1.4$,
where the disks become more spread out, $P_{6}$ drops again.
At $\phi = 0.71$ in Fig.~\ref{fig:7}(e),
for low drives $P_{6} \approx 0.55$, and then $P_6$ gradually increases with increasing
drive up to a value of $P_6=0.9$,
indicating that most of the sample has developed triangular ordering.
Finally, for $\phi = 0.85$ in Fig.~\ref{fig:7}(f), at the lowest drives the system forms
a polycrystalline solid containing a small number of defects, so that the initial
value of $P_6 \approx 0.81$, while as
$F_{D}$ increases,
the polycrystal anneals into a single domain crystal that is aligned in the direction of drive,
with $P_{6} = 0.99$, indicating almost complete triangular ordering.

For $0.3 < \phi <  0.85$, the $P_6$ curves in Fig.~\ref{fig:7} show
a small peak
near $F_{D}/F_p = 0.2$  due to the pile up or clustering effect.
Within the clusters the local density $\phi_{\rm loc}$
is $\phi_{\rm loc}\approx 0.85$, producing increased sixfold ordering and a
corresponding increase in $P_6$.
Once the drive is large enough to break apart these clusters,
there is a drop in $P_{6}$ as the system enters the
homogeneous moving phase.

\begin{figure}
  \includegraphics[width=3.5in]{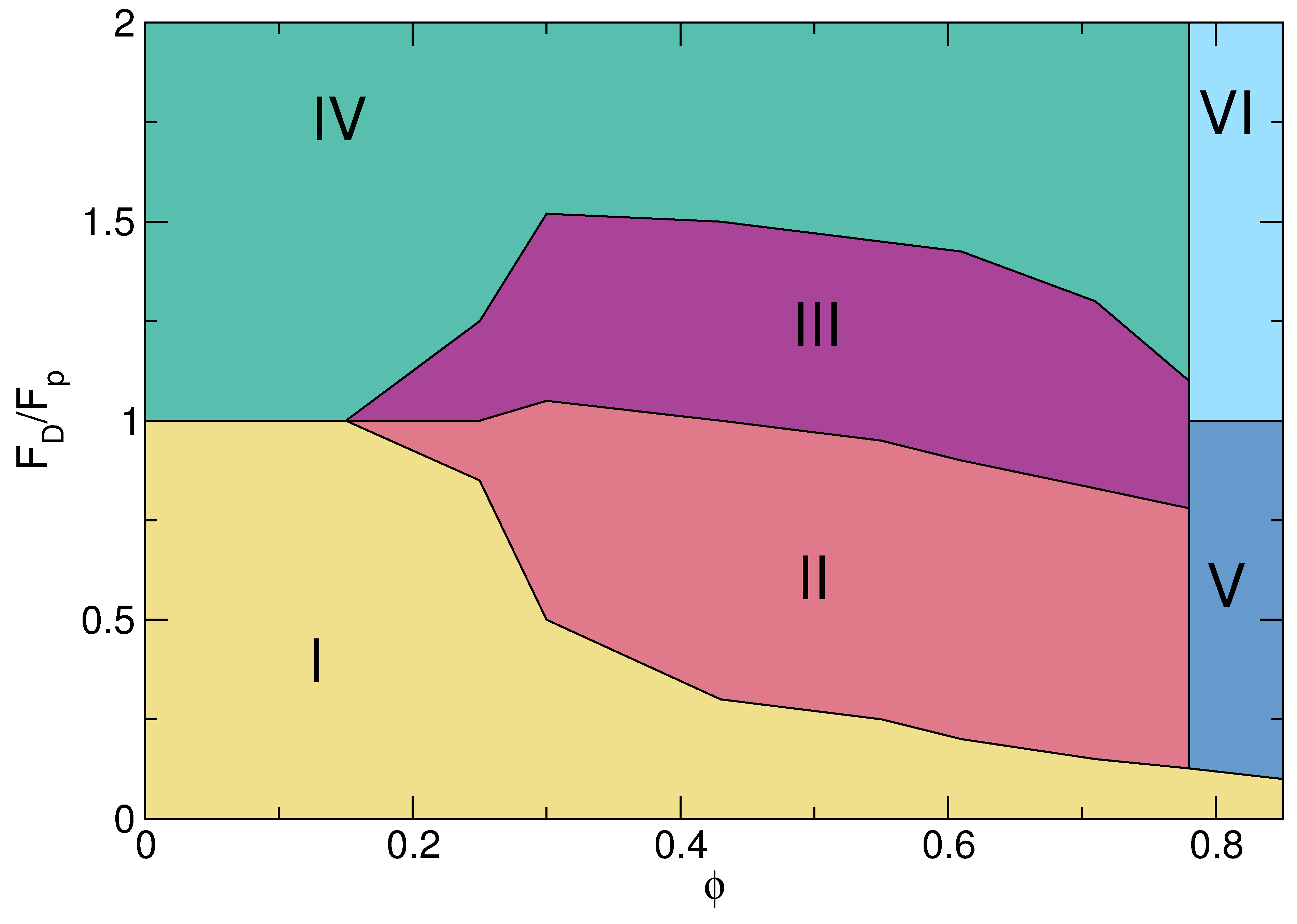}
\caption{ Schematic phase diagram as a function of $F_{D}/F_p$ vs
  $\phi$ for the system in Fig.~\ref{fig:1}.
  I: Pinned or clogged state.
  II: Homogeneous plastic flow.
  III: Density phase separated state.
  IV: Moving smectic or moving chain state.
  V: Moving polycrystalline state.
  VI: Moving crystal state.
}
\label{fig:8}
\end{figure}

From the features in the velocity-force curves, $P_{6}$, $\langle \delta y^2\rangle$,
and the disk configurations,
we can construct a schematic phase diagram of the evolution of the different phases,
as shown in Fig.~\ref{fig:8}.
Phase I corresponds to the pinned or clogged state,
phase II is homogeneous disordered plastic flow,
phase III is the density phase separated state,
phase IV is the moving smectic or moving chain state,
phase V is the moving polycrystalline state, and
phase VI is the moving single domain crystal state.
At low $\phi$ where few disk-disk collisions occur,
the system passes directly from a pinned state for $F_{D}/F_p < 1.0$
to a moving state that is similar
to the moving chain state illustrated in Fig.~\ref{fig:5}.
As $\phi$ increases, the depinning threshold drops due to collisions between
unpinned and pinned disks,
and above depinning
the system enters a homogeneous plastic flow state similar to that
shown in Fig.~\ref{fig:2}(c,d).
At higher drives, the density phase separated state illustrated in
Fig.~\ref{fig:3}(a) forms, followed by a transition to the
moving chain state shown in Fig.~\ref{fig:3}(d).
For $\phi>0.77$, 
the disks depin into a moving
polycrystalline state which transitions
to a single moving crystal at high drives.
We expect that for $\phi \geq 0.9$, the system would form
a crystalline pinned or jammed state that would depin elastically into
a moving single crystal state.

\section{Varied Pinning Density}

\begin{figure}
  \includegraphics[width=3.5in]{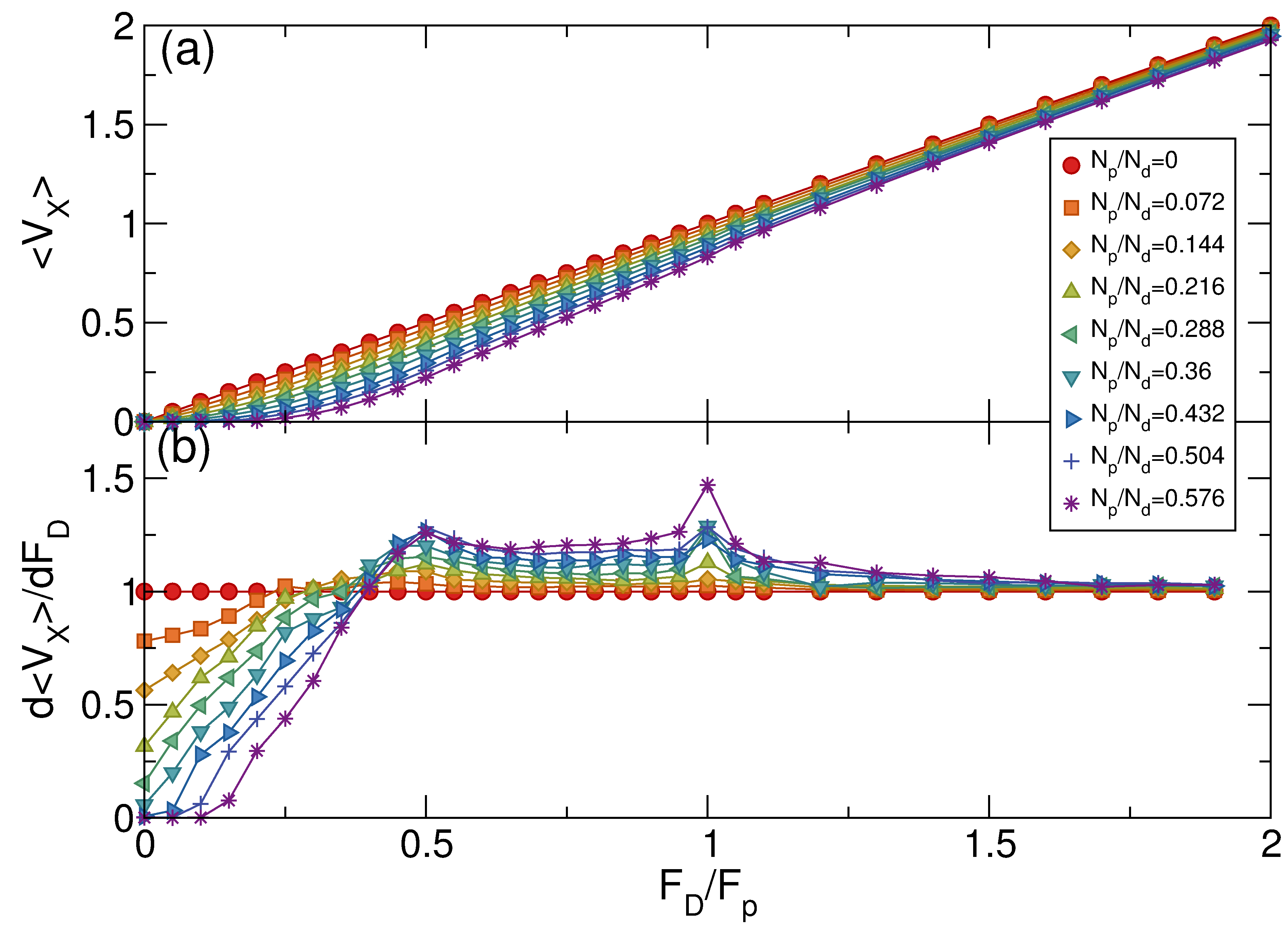}
\caption{ (a) $\langle V_{x}\rangle$ vs $F_{D}/F_{p}$ at $\phi = 0.55$
  and $F_{p} = 1.0$ for
  $N_{p}/N_{d} = 0.0$, 0.072, 0.144, 0.216, 0.288, 0.36,  0.432, 0.504,
  and $0.576$, from top to bottom.
  (b) The corresponding
  $d\langle V_{x}\rangle/dF_{D}$ vs $F_D/F_p$ curves showing peaks near
  $F_D/F_p=0.5$ and $F_D/F_p=1.0$.
}
\label{fig:9}
\end{figure}

\begin{figure}
  \includegraphics[width=3.5in]{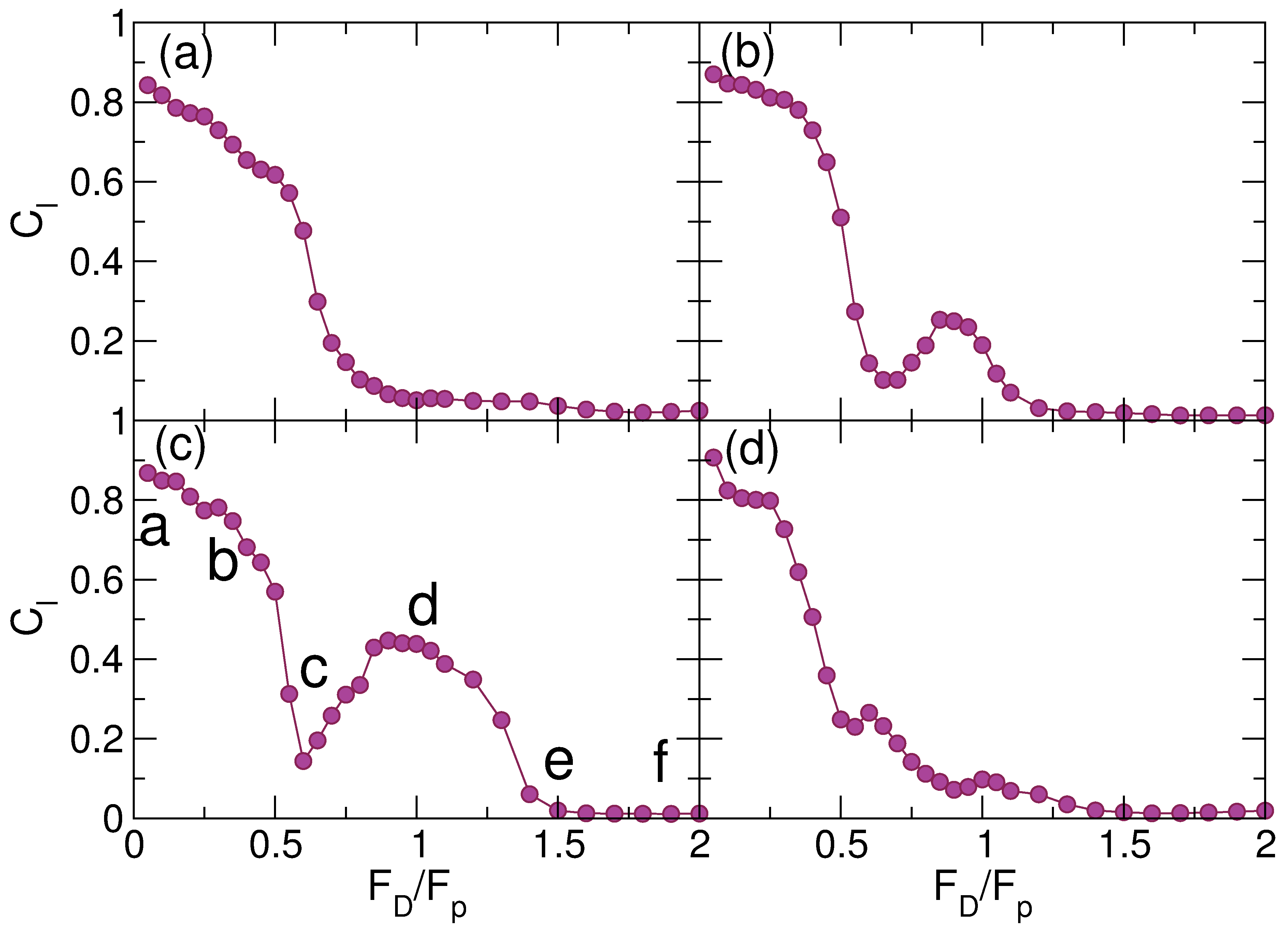}
\caption{ Cluster size $C_{l}$ vs $F_{D}/F_p$ for the system in Fig.~\ref{fig:9} at
$N_{p}/N_{d} =$ (a) 0.072, (b) $0.216$, (c) $0.288$, and (d) $0.432$. The local peaks
  in (b) and (c) correspond to the formation of a density phase separated state.
  In panel (c) the lettering indicates the $F_{D}/F_p$ values represented in the
  real space images in Fig.~\ref{fig:11}.
}
\label{fig:10}
\end{figure}

We next consider the case of a fixed disk density of $\phi = 0.55$, where
$N_{d} =  2500$, and vary the number of pinning sites to give a ratio
of $N_{p}/N_{d}$ ranging from
$N_p/N_d= 0$ to $N_p/N_d=0.576$.
In Fig.~\ref{fig:9}(a,b) we show
$\langle V_{x}\rangle$ and $d\langle V_{x}\rangle/dF_{D}$ versus $F_{D}/F_{p}$
for a sample with $F_{p} = 1.0$.
There is one peak in $d\langle V_{x}\rangle/dF_{D}$
near $F_{D}/F_{p} = 1.0$,  the drive above which all of the disks are moving,
and a second peak near $F_{D}/F_p = 0.5$,
the drive at which the clogged state breaks apart.
We observe a similar set of phases
as described above, but find that
the density phase separated state is more prominent at lower pinning density,
as shown in the plots of $C_{l}$ versus $F_{D}/F_{p}$
in Fig.~\ref{fig:10} for $N_{p}/N_{d} = 0.072$,  0.216, 0.288, and $0.432$.
In particular, 
$N_{p}/N_{d} = 0.216$ in Fig.~\ref{fig:10}(b) and
$N_p/N_d=0.288$ in Fig.~\ref{fig:10}(c) exhibit strong
peak features associated with the density phase separated state.

\begin{figure}
  \includegraphics[width=3.5in]{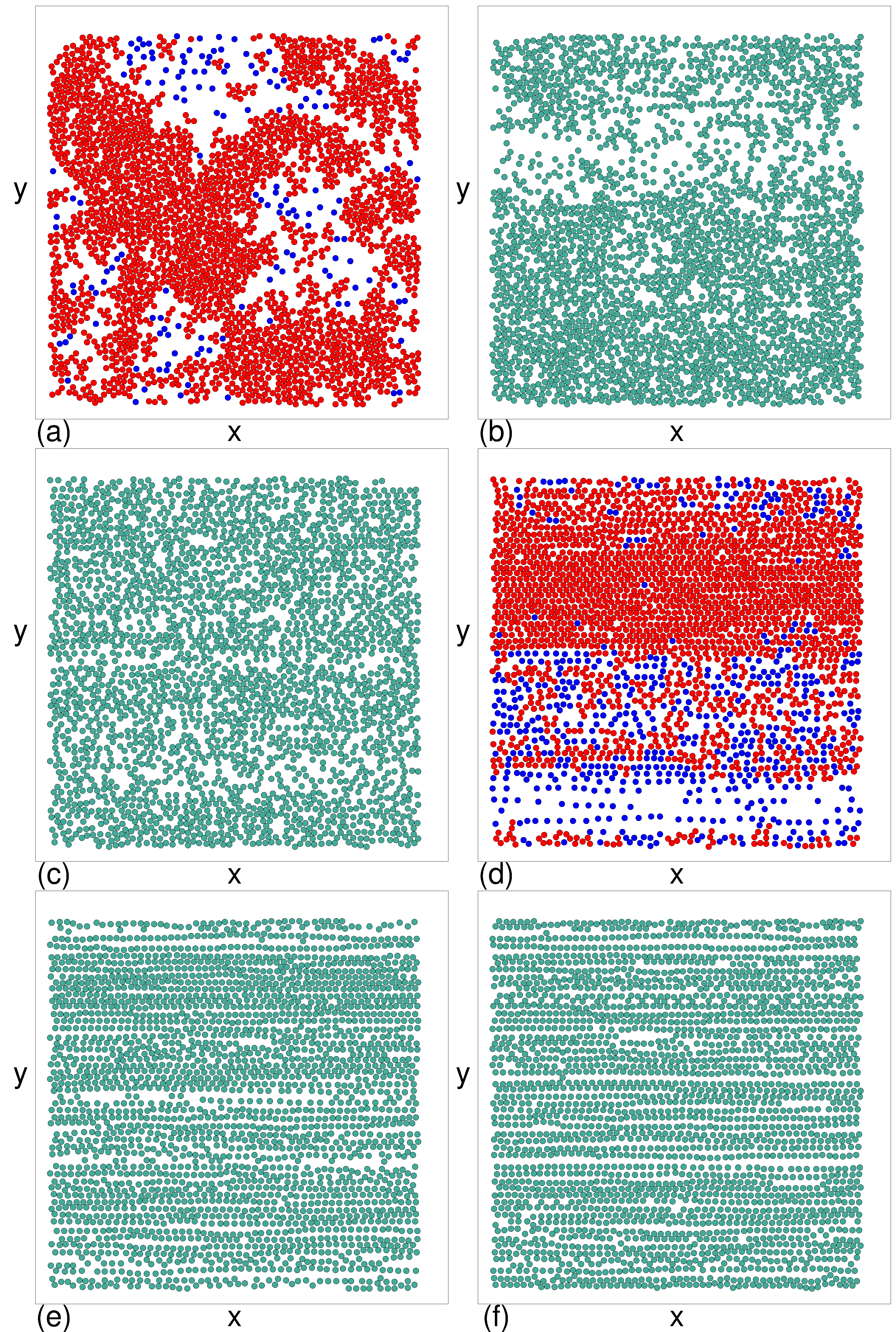}
\caption{The disk positions for the system in
  Figs.~\ref{fig:9} and \ref{fig:10} at
  $N_p/N_d = 0.288$ for drive values marked with letters in Fig.~\ref{fig:10}(c).
  In panels (a) and (d),
  red disks are part of clusters containing three or more disks, while blue disks
  are isolated or in a cluster containing only two disks.
  (a) The pinned cluster state at $F_{D}/F_p = 0.05$.
  (b) At $F_{D}/F_p = 0.3$ the moving disks form more spread out clusters.
  (c) At $F_{D}/F_p = 0.6$, corresponding to the local minimum of $C_{l}$
  in Fig.~\ref{fig:10}(c), a homogeneous disordered state forms.
  (d) At $F_{D}/F_p = 1.05$, corresponding to the peak in $C_{l}$ in Fig.~\ref{fig:10}(c),
  a density phase separated state forms.
At (e) $F_{D}/F_p = 1.5$ and (f) $F_{D}/F_p = 2.0$, the disks are in a moving chain state.
}
\label{fig:11}
\end{figure}

To show more clearly the evolution of the cluster state,
in Fig.~\ref{fig:11} we illustrate the disk positions for the
system at $N_{p}/N_{d} = 0.288$ for increasing $F_{D}$.
The letters a through f in Fig.~\ref{fig:10}(c)
indicate the values of $F_{D}/F_p$ that match the images.
In Fig.~\ref{fig:11}(a) at $F_{D}/F_p = 0.05$, where $C_{l} = 0.85$,
the system forms a clogged state.
Within the cluster regions, which are colored red,
the disk density is close to $\phi = 0.85$,
and these clusters are separated by low density regions of disks.
As the drive increases, the
large cluster becomes more spread out,
as shown in Fig.~\ref{fig:11}(b) for $F_{D}/F_p = 0.3$, where $C_{l}$ drops to
$C_l=0.78$.
At $F_{D}/F_p = 0.6$ in Fig.~\ref{fig:11}(c), which corresponds to a local
minimum in $C_{l}$ in Fig.~\ref{fig:10}(c),  the disks
are completely spread out and form a homogeneous disordered phase.
In Fig.~\ref{fig:11}(d) at $F_{D}/F_p = 1.05$,  which corresponds to a local
maximum in $C_{l}$ in Fig.~\ref{fig:10}(c),
a density phase separated state appears.
Disks that are in a cluster containing at least three disks are colored red in order to more
clearly highlight the dense region, within which the disks have developed triangular
ordering.
As the drive is further increased, the disks
spread apart in the direction transverse to the drive to form
the moving chain state
illustrated in Fig.~\ref{fig:11}(e,f)
at $F_{D}/F_p = 1.5$ and $F_D/F_p=2.0$, respectively,
which also coincides with
a reduction of $C_{l}$ in Fig.~\ref{fig:10}(c).
For $N_p/N_d = 0.55$ and above, the density phase
separated state becomes less well defined,
as indicated in Fig.~\ref{fig:10}(d) at $N_{p}/N_{d} = 0.432$.

\begin{figure}
  \includegraphics[width=3.5in]{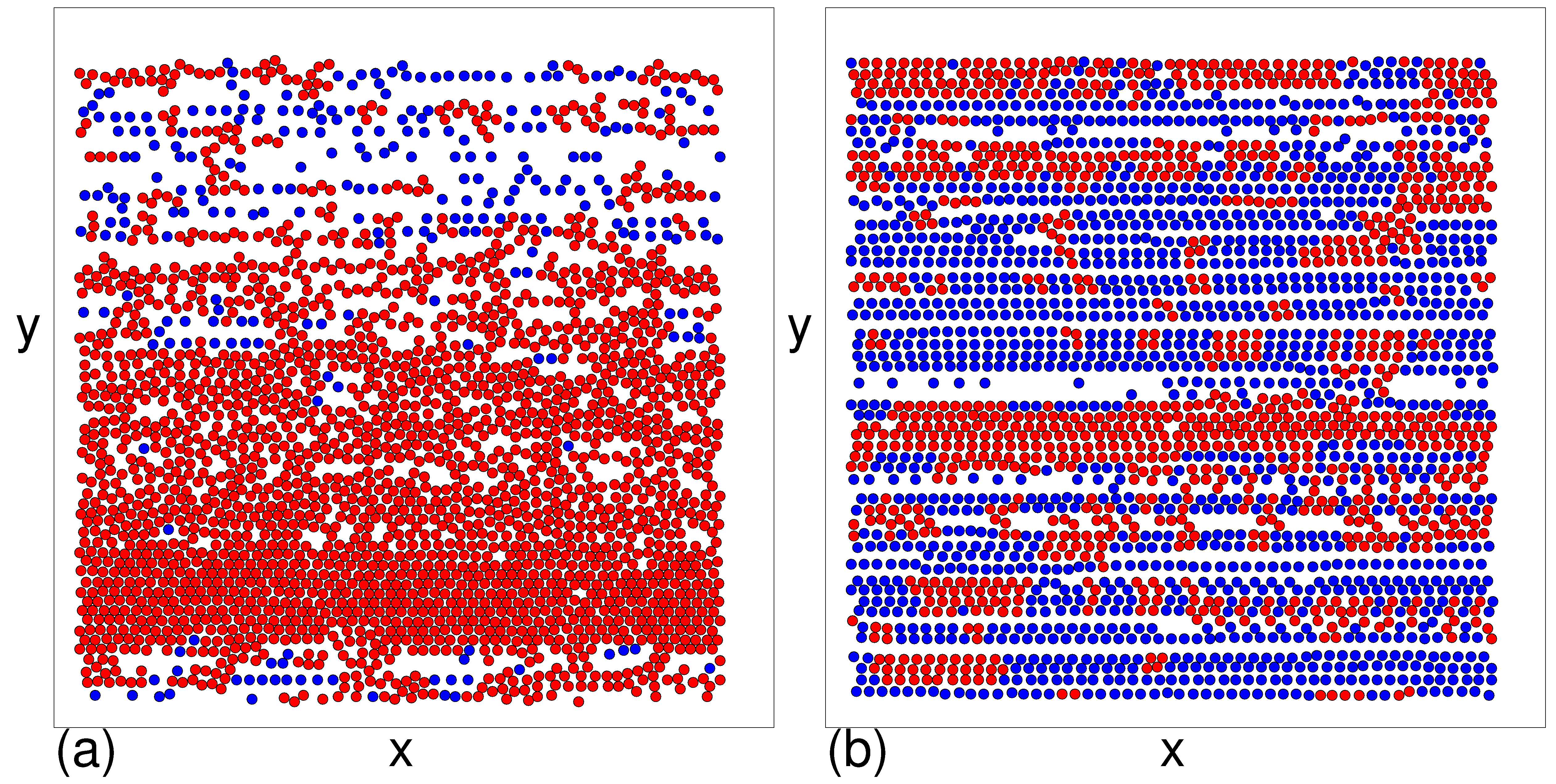}
\caption{ The disk positions for a system with
  $\phi = 0.55$ at $N_{p}/N_{d} = 0.072$, where there is no peak in
  $C_{l}$ in Fig.~\ref{fig:10}(a).
  Red disks are part of clusters containing three or more disks, while blue disks
  are isolated or in a cluster containing only two disks.
  (a) A density phase separated state at $F_{D}/F_p = 0.3$.
  (b) A moving chain state forms at higher drives, shown here at
  $F_{D}/F_p = 1.5$.
}
\label{fig:12}
\end{figure}

\begin{figure}
  \includegraphics[width=3.5in]{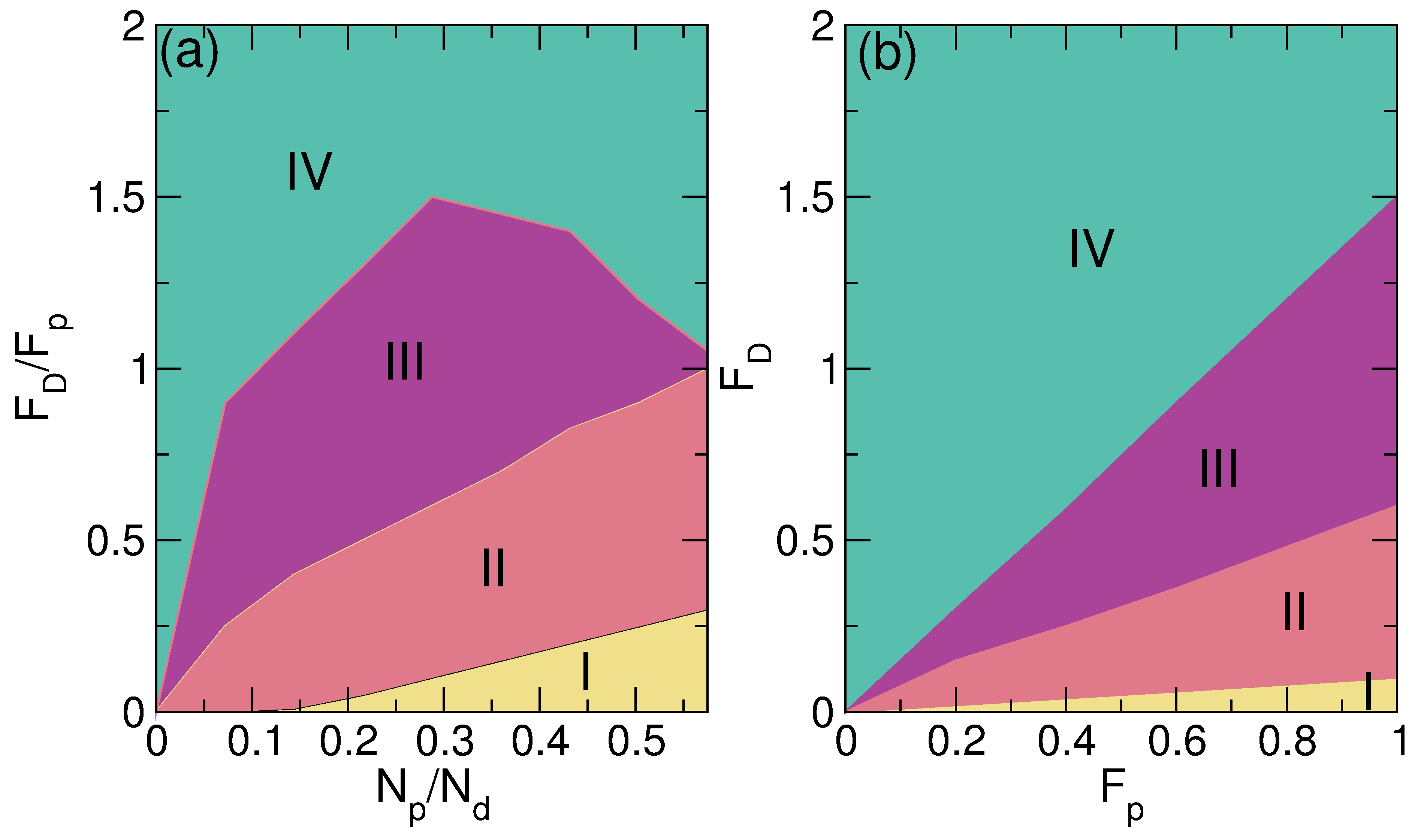}
\caption{ (a) Schematic phase diagram 
  as a function of $F_{D}/F_p$ vs $N_p/N_d$ for the system in
  Fig.~\ref{fig:9} at fixed $\phi=0.55$.
  I: Pinned or clogged state.
  II: Homogeneous plastic flow.
  III: Density phase separated state.
  IV: Moving smectic or moving chain state.
  (b) Phase diagram for the same system at $\phi=0.55$ and $N_p/N_d=0.288$
  as a function of $F_D$ vs $F_p$.
}
\label{fig:13}
\end{figure}

In Fig.~\ref{fig:10}(a) at $N_p/N_d=0.072$, 
although $C_{l}$ does not show a peak near $F_{D}/F_p = 1.0$,
there is still
a pronounced density phase separated state;
however, this phase has shifted to lower $F_{D}/F_p$.
Since the low density clogged state transitions directly into the
flowing density phase separated state, 
there is no dip in $C_l$.
The density phase separated state breaks apart at lower
values of $F_{D}/F_{p}$ compared to samples with higher values of $N_p/N_d$.
In Fig.~\ref{fig:12}(a) we show the disk
configurations at $N_{p}/N_{d} = 0.072$ and $F_{D}/F_p = 0.3$ where
a density phase separated state appears,
while in Fig.~\ref{fig:12}(b) we illustrate the moving chain phase that forms
at $F_D/F_p=1.5$ in the same system.
From the images we can construct a schematic phase diagram for the
$\phi = 0.55$ sample as a function of
$F_{D}/F_p$ versus $N_{p}/N_{d}$, as shown in Fig.~\ref{fig:13}(a), which
highlights the extents of regions I through IV.
Here, the widths of regions I and II
grow with increasing $N_{p}/N_d$, while
region III reaches its largest extent near $N_{p}/N_{d} = 0.3$.
We note that for $N_{p}/N_{d}  = 0$,  the system forms a moving disordered state
for all $F_D>0$.

\begin{figure}
  \includegraphics[width=3.5in]{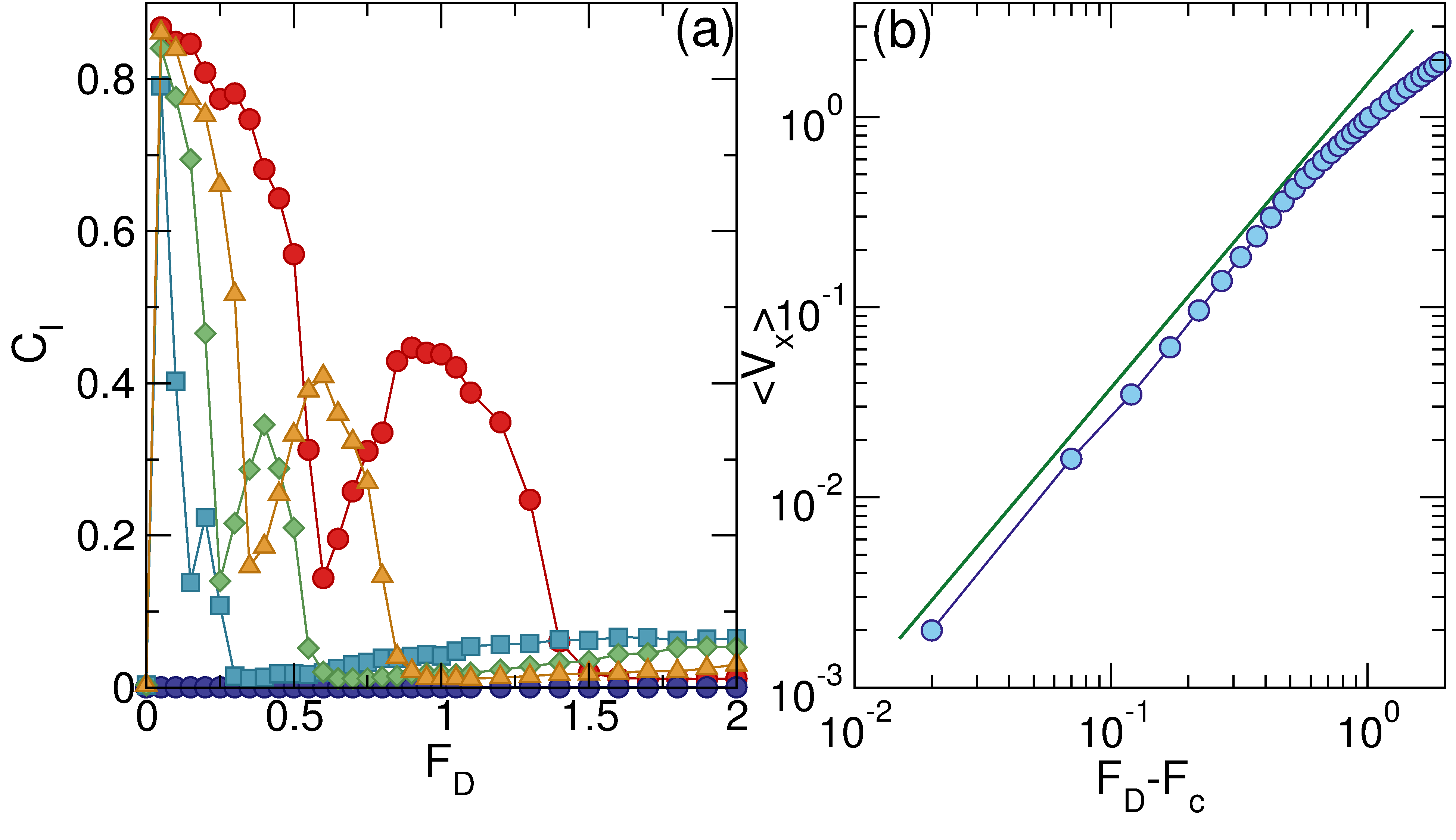}
\caption{
  (a) The cluster size $C_{l}$ vs $F_{D}$ for
  samples with $\phi = 0.55$ and $N_{p}/N_{d} = 0.288$
  at
  $F_{p} = 0.0$ (blue circles), $0.2$ (blue squares),
  $0.4$ (green diamonds), $0.6$ (orange triangles),
  and $1.0$ (red circles).
  (b) $\langle V_x\rangle$ vs
  $F_{D} - F_{c}$ for
  the velocity-force curve obtained at $\phi = 0.55$ and $N_{p}/N_{D} = 0.576$. The solid
line is a power law fit with an exponent of  $\beta=1.6$.
}
\label{fig:14}
\end{figure}

We have considered varying $F_p$ while holding
$\phi$ and $N_{p}/N_d$ fixed,
and find that the same general phases appear.
The onset of the phases shifts linearly
to higher values of $F_{D}$ with increasing $F_{p}$, as shown in
the phase diagram of $F_D$ versus $F_p$ in Fig.~\ref{fig:13}(b).
In Fig.~\ref{fig:14}(a) we plot $C_{l}$ versus
$F_{D}$ in a sample with $\phi = 0.55$ and $N_{p}/N_{d} = 0.288$ for
$F_{p} = 0.0$, $0.2$, $0.4$, $0.6$, and $1.0$.  The width of the
peak in $C_l$ associated with density phase separation shifts to higher values
of $F_D$ with increasing $F_p$.

For depinning in systems with longer range interactions,
such as superconducting vortices, colloidal particles, and electron crystals,
scaling near the depinning threshold has been observed
in the velocity-force curves,
which have the form $V \propto (F_{D} - F_{c})^{-\beta}$.
In plastic depinning, where particles exchange neighbors as they move,
$\beta > 1.0$,
while for elastic depinning, in which the particles maintain the same neighbors as
they move, $\beta<1.0$ \cite{1}.
In systems
with long range Coulomb interactions \cite{6} and
screened Coulomb interactions \cite{10,35}, plastic depinning is
associated with exponents of
$\beta \approx 1.65$ and $\beta \approx 2.0$, respectively.
More recently, simulations of depinning of
superconducting vortices with a Bessel function
vortex-vortex interaction give $\beta = 1.3$ \cite{36}.
Thus, it is interesting to ask whether
similar scaling of the velocity-force curves occurs in the disk system.
In Fig.~\ref{fig:14}(b) we plot $\langle V_{x}\rangle$ versus
$F_{D} - F_{c}$ on a log-log scale for
a sample with $\phi = 0.55$ at $N_{p}/N_{D} = 0.576$.
The solid line indicates a scaling fit with
$\beta = 1.6$.
At higher drives, well above depinning, the slope of the velocity-force curve
becomes linear, as expected since the effectiveness of the pinning is lost in this regime.
In general, we find that for $N_{p}/N_{d} > 0.288$,
the velocity-force curves can be fit to a power law with $1.4 < \beta < 1.7$.
For $N_{p}/N_{d} < 0.288$, the depinning threshold $F_c=0$
since there are few enough pinning sites that some disks
can pass completely through the system without being trapped directly by pinning
or indirectly by becoming lodged behind pinned disks.
This result indicates that scaling near the depinning threshold for plastic
flow is robust in systems with short range repulsion as well as
in those with longer range repulsion.

\section{Discussion}
The dynamic density phase separation we find has not been
observed in studies of superconducting vortices or colloids driven over random disorder.
There have
been cases where a phase separation into high
density regions as well as stripe ordering occurs for particles
driven over random disorder when the pairwise interactions between
particles include
both a repulsive and an attractive term \cite{37,38};
however, in the disk system we consider here, the disk-disk interaction is
purely repulsive.
The phase separation we observe can be
viewed as an example of active matter clustering of the type found in
simulations
of hard disks undergoing active Brownian motion or run-and-tumble type dynamics.
In the active matter systems, when the
activity is high enough, the particles phase separate into
a dense solidlike region and a low density fluid \cite{29,30,31,32} 
due to a combination of the nonequilibrium nature of the fluctuations
and the fact that the mobility of the particles is dependent on the local
particle density \cite{30}.
In the driven disk system, velocity fluctuations transverse to the driving direction are
generally largest when there is a coexistence of disks being pinned or slowed down
by the pinning along with faster moving unpinned disks.
When the disks collide with each other,
they generate velocity fluctuations that have
a ballistic component in the transverse direction, similar to the
motion of active particles.
This also produces time intervals in the transverse diffusion that exhibit superdiffusive
behavior similar to that found in active matter systems \cite{32}.
Additionally, the disks have a reduced mobility
when the disk density increases.
When the drive is large enough,
both the speed differential of the disks and the velocity fluctuations transverse to the
drive are lost, and since these effects are necessary to produce the clustering and the
density phase separation, the clustering and density phase separation also disappear.
The same effects could arise in systems with longer range interactions;
however, the large energy cost of high density regions would
suppress the density phase separation we observe for the short
range repulsive disks.
Experimentally, the dynamic phase separation could be
observed using colloids that have only steric interactions
moving over random substrate.
Experiments with quasi-2D granular systems could include grains
flowing over a rough landscape under the influence of gravity or shaking.
In our work we focus on the case of monodisperse disks,
so that the system forms triangular ordering in the dense phase.
We expect that many of the general results should remain robust in
nonmonodisperse systems; however, the phases may shift to lower
densities since it is known that bidisperse disk systems
exhibit jamming at a lower density than monodisperse disks \cite{N}.
In addition, the
bidisperse disks would have an amorphous structure in the dense phase separated
regions.
There could be additional behavior that arises in
nonmonodisperse systems since the flow could induce
a phase separation of the different species of disks, as well as a density phase
separation.

\section{Summary}
We have numerically examined the dynamical phases for monodisperse
repulsive disks driven
over random disorder. Despite the simplicity of this system,
we observe a rich variety of distinct dynamics,
many of which have significant differences from the dynamic phases
observed for other systems of
collectively interacting particles with longer range repulsion,
such as vortices in type-II superconductors and colloids with Yukawa interactions.
The phases we find include a heterogeneous clogged state
where the disks form local immobile clumps,
a homogeneous disordered plastic flow state, a moving density phase separated state
where the system forms a dense region with mostly triangular ordering coexisting with a
low density disordered phase,
and a stripe or chainlike state at higher drives.
The density phase separation occurs due to the density dependent mobility of the
disks
and the short range nature of their interaction with each other, which permits the
disks to  pack closely together with little overlap energy.
In contrast,
in vortices, Coulomb systems, and charged colloids,
the longer range repulsion would prevent density phase separated states from
forming since
more homogeneous states would have a much lower
particle-particle interaction energy.
From the features in the transverse diffusion, structure factor, and velocity-force curves, we
map the evolution of the different phases as a function of disk density,
pinning site density, and pinning force.
Our results suggest that the dynamic density phase separation and the chainlike state
should be general features in systems
with short range steric interactions driven over random disorder.
These effects could be observed experimentally using
sterically interacting colloids or granular matter flowing over random disorder.

\acknowledgments
This work was carried out under the auspices of the 
NNSA of the 
U.S. DoE
at 
LANL
under Contract No.
DE-AC52-06NA25396.

\end{document}